\documentclass[apl, twocolumn ,amsmath,amssymb,reprint]{revtex4}
\usepackage{graphicx}
\usepackage{epsfig} 
\usepackage{color}
\usepackage{epstopdf}
\usepackage{bm}
\newcommand{\figurewidth}{3.4in}
\newcommand{\FG}[1]{Fig.~\ref{#1}}
\newcommand{\EQ}[1]{Eq.~(\ref{#1})}

\newcommand{\bra}[1]{{\langle\, #1 \,\vert\,}}

\newcommand{\e}[1]{{\rm e}^{#1}}
\newcommand{\difr}{\;{\rm d}^3 {\bm r}}
\newcommand{\basis}{{\bm \tau}_{\alpha n}}
\newcommand{\argg}{({\bm G})}

\renewcommand{\Re}{\operatorname{Re}}
\renewcommand{\Im}{\operatorname{Im}}

\begin{document}
\title{Effective Passivant Pseudopotentials for semiconductors: beyond the spherical approximation}

\date{\today}
\author{J. R. C\'ardenas}
\email[Email: ]{jairocardenas@mail.uniatlantico.edu.co}
\affiliation{Programa de F\'isica, Universidad del Atl\'antico, Km 7 antigua v\'ia a Puerto Colombia, Barranquilla, Colombia}

\begin{abstract}
An effective atomic pseudopotential for passivation of semiconductor surfaces is presented. It is shown that the spherical approximation used in the effective and  empirical pseudopotential methods is not suitable for describing passivants and that, instead, they have to be regarded as complex quantities. Since the new pseudopotentials cannot be handled as usual bulk ones, the way of implementing them in atomistic methods is described here, together with a methodology for extracting them trough an analytic connection to density functional theory, meaning that these surface potential keeps the {\it ab inito} identity. The effectiveness and high transferability of the approach is demonstrated by generating passivant potentials for six different semiconductor compounds (GaAs, AlAs, Si, Ge, CdSe and ZnO) and testing them on different kinds of surfaces, including a passivated 68 atoms CdSe quantum wire in wurtzite structure, obtaining always an excellent agreement to density functional 
theory calculations done on the same 
systems.
\end{abstract}

\maketitle

The fabrication of semiconductor nanostructures is nowadays controlled at the atomic scale. At the same time, new experimental ideas make evident the relevance of the atomic identity of the experimental samples \cite{yakunin2007,Gamalski2016,Kudelski2007,Soumitra2006,Mokari1787,Besombes2004,yoffe2001}. In this sense, to be able to theoretically study low dimensional semiconductor structures, it is equally required to apply modern atomistic methods, capable of considering characteristics such as precise composition, shape and size of the structures \cite{Huaxiang1997,bester2009,wang2004,Zirkelbach2015,Luo2010,puangmali2008}. 

One of the major inconveniences of going to the atomic limit with theoretical methods, as far as semiconductor nanostructures are concerned, are the number of atoms conforming the samples. When we talk about semiconductor quantum wells, wires or dots, we are talking about systems composed of several thousands atoms. These sizes, still in the nanometer regime, are far beyond the limit of several {\it ab initio} techniques such as density functional theory (DFT) in terms of computational costs \cite{bester2009}. 

Different ways to try to overcome the computational demands of calculations of nanostructures with high number of atoms have been proposed. Specialized basis sets, for example, have been effectively used to calculate band structures and total energies of solids \cite{Mattheiss1986,PhysRevB.59.15806}. Other alternative is the use of optimized atomic pseudopotentials (PSP), where the general idea is to soften the innermost part of the potentials and all-electron wavefunctions by well behaved functions leading to a faster plane-wave convergence \cite{Hamann1979,rappe1990,Phillips1958,Cohen1966,Chelikowsky1976,Ito1990,cardenas2012,Fritsch2006}.   

Based on a fitting procedure, the empirical pseudopotential method (EPM) was first used to construct atomic pseudopotentials, where they were adjusted empirically until reproducing the experimentally determined energy levels of bulk crystals \cite{Phillips1958,Cohen1966,Chelikowsky1976}. This idea has been reviewed different times \cite{wang1995,fu1997}, until arriving to the generation of the atomic effective pseudopotentials (AEPs)  \cite{cardenas2012,Zirkelbach2015}, whose construction is purely based on DFT calculations, is free of fitting procedures, and provide pseudopotentials with high transferability.  

Spherical pseudopotentials have been widely used in semiconductor physics \cite{Saurabh2014,Ming2011,Shokeen2010,Fritsch2006,Jiseok2011}, but an extra difficulty arises in the study of structures where surfaces play a significant role, such as colloidal systems, since the surface potentials behave differently than in the bulk. This area of research is currently very active and empirical spherical PSPs have been used \cite{puangmali2008,Molina2012,Wang1996,Hu2002}, but their accuracy has not been formally demonstrating so far \cite{Jiseok2011}. 

In this paper, an effective pseudopotential for passivation is presented. It is shown that in order to effectively passivate semiconductor surfaces, it is mandatory to go beyond the spherical approximation and, hence, consider a full complex character of the passivant potentials in reciprocal space. A way to extract both, the imaginary and real components, is presented, based on an analytic expressions that relates the AEPs and the effective crystal potential from DFT. In this way, if the bulk AEPs are previously known \cite{cardenas2012}, the crystal potential can be reconstructed and from this extract effective passivant potentials (EPPs).

It has to be mentioned that the methodology introduced here to generate EPPs is completely general, meaning that the use of non-spherical pseudopotentials for the correct description of semiconductor surfaces holds for any method (EPM,AEPs, etc.). 

The starting point is the local self-consistent effective potential delivered by DFT, which can be written as a sum of atomic potentials centered on the atoms \cite{cardenas2012,Zirkelbach2015}
\begin{equation}
V_{\rm loc}({\bm r}) =  \sum_\alpha^{N_{\rm species}}  \sum_{n}^{N_\alpha} v_\alpha(\bm r - \basis)    , \label{vloc0}
\end{equation}
where $\alpha$ describes the atom type and runs from one to the number of atomic species $N_{\rm species}$ and $n$ labels each one of the $N_\alpha$ atoms of type $\alpha$. The atomic positions are given by $\basis$. 

By Fourier transforming both sides of the last equation, the local effective potential transforms into a sum of potentials centered at the origin in reciprocal space
\begin{equation}
V_{\rm loc} \argg = \sum_\alpha^{N_{\rm species}}  \sum_{n}^{N_\alpha} 
 \e{-i {\bm G} \cdot {\bm \tau}_{\alpha n} } v_\alpha(\bm G) , \label{eq:vlocstandard}
 \end{equation}
 with
 \begin{equation}
 v_\alpha(\bm G) = \frac{1}{\Omega_c}\int_{\infty} v_\alpha(\bm r)\e{-i {\bm G} \cdot  {\bm r}}  \difr   ,
\end{equation}
and
\begin{equation}
V_{\rm loc} \argg = \frac{1}{\Omega_c}\int_{\Omega_c} V_{\rm loc}({\bm r})\e{-i {\bm G} \cdot  {\bm r}} , \label{eq:vloc_r}
\end{equation}
where $V_{\rm loc} \argg$ is known directly from DFT (or even reconstructed from experimental data as in the EPM). 

Expanding explicitly the right hand side of \EQ{eq:vlocstandard} in terms of the atom types including the passivants:
\begin{equation} \label{eq:vloc1}
\begin{split}
V_{\rm loc}^{(1)} \argg  &= \sum_i^{N_{H_a}}e^{i{\bm G}\cdotp{\bm \tau}_i^{(1)}}v_{H_a}({\bm G}) +  \sum_j^{N_{H_c}}e^{i{\bm G}\cdotp{\bm \tau}_j^{(1)}}v_{H_c}({\bm G}) \\ 
 & + \sum_n^{N_a^{(1)}}e^{i{\bm G}\cdotp{\bm \tau}_n^{(1)}}v_a({\bm G}) + \sum_m^{N_c^{(1)}}e^{i{\bm G}\cdotp{\bm \tau}_m^{(1)}}v_c({\bm G})   ,
\end{split}
\end{equation}
where $v_{a} (v_{c})$ stands for the anion (cation) potential and $v_{H_{a(c)}}$ its passivant. 

If the bulk (anion and cation) potentials are previously known  \cite{cardenas2012}, the last two therms in \EQ{eq:vloc1} can be determined, and the equation solved for the passivant potentials
\begin{equation}
  \sum_i^{N_{H_a}}e^{i{\bm G}\cdotp{\bm \tau}_i^{(1)}}v_{H_a}({\bm G}) +  \sum_j^{N_{H_c}}e^{i{\bm G}\cdotp{\bm \tau}_j^{(1)}}v_{H_c}({\bm G}) = \gamma^{(1)}({\bm \tau},{\bm G}), \label{eq:eq1}
\end{equation}
where
\begin{equation}
\gamma^{(1)}({\bm \tau},{\bm G}) = V_{\rm loc}^{(1)} \argg - \sum_n^{N_a^{(1)}}e^{i{\bm G}\cdotp{\bm \tau}_n^{(1)}}v_a({\bm G}) - \sum_m^{N_c^{(1)}}e^{i{\bm G}\cdotp{\bm \tau}_m^{(1)}}v_c({\bm G}). \label{eq:gamma1}
\end{equation}

By keeping the same composition and supercell dimensions, but changing the effective size of the material slab, we can write the same expresion for a second system
\begin{equation}
  \sum_i^{N_{H_a}}e^{i{\bm G}\cdotp{\bm \tau}_i^{(2)}}v_{H_a}({\bm G}) +  \sum_j^{N_{H_c}}e^{i{\bm G}\cdotp{\bm \tau}_j^{(1)}}v_{H_c}({\bm G}) = \gamma^{(2)}({\bm \tau},G), \label{eq:eq2}
\end{equation}
where
\begin{equation}
\gamma^{(2)}({\bm \tau},{\bm G}) = V_{\rm loc}^{(2)} \argg - \sum_n^{N_a^{(2)}}e^{i{\bm G}\cdotp{\bm \tau}_n^{(2)}}v_a({\bm G}) - \sum_m^{N_c^{(2)}}e^{i{\bm G}\cdotp{\bm \tau}_m^{(1)}}v_c({\bm G}). \label{eq:gamma2}
\end{equation}

Notice that the only restriction between \EQ{eq:eq1} and \EQ{eq:eq2} is the size of the supercell, since this leads to the same ${\bm G}$-point grid in both cases and makes easier and more precise the numerical calculations latter. 

The bulk potentials are considered to fulfill the spherical approximation, meaning that they are real quantities \cite{cardenas2012,wang1995,fu1997,Phillips1958,Cohen1966,Chelikowsky1976}, but $V_{\rm loc}^{(i)}$ as well as the summations in \EQ{eq:gamma1} and \EQ{eq:gamma2} are complex, therefore; the $\gamma^{(i)}({\bm \tau},{\bm G})$ functions are complex too.

The real and imaginary components of the passivant potentials $v_{H_\alpha}$ can be extracted from \EQ{eq:eq1} and \EQ{eq:eq2}, by solving the system of four equations (two from the real and two from the imaginary parts), which in matricial form reads: 
\begin{equation}
 \begin{pmatrix}
 \beta_{a}^{(1)}  &  \alpha_{a}^{(1)} & \beta_{c}^{(1)}  & -\alpha_{c}^{(1)} \\
 \alpha_{a}^{(1)} & -\beta_{a}^{(1)}  & \alpha_{c}^{(1)} & \beta_{c}^{(1)}   \\
 \beta_{a}^{(2)}  &  \alpha_{a}^{(2)} & \beta_{c}^{(2)}  & -\alpha_{c}^{(2)} \\
 \alpha_{a}^{(2)} & -\beta_{a}^{(2)}  & \alpha_{c}^{(2)} & \beta_{c}^{(2)}   
\end{pmatrix}
 \begin{pmatrix}
 \Re[v_{H_a}] \\ 
 
 \Im[v_{H_a}] \\ 
 
 \Re[v_{H_c}] \\ 
 
 \Im[v_{H_c}] 
\end{pmatrix}
=
 \begin{pmatrix}
 \Re[\gamma^{(1)}] \\ \Im[\gamma^{(1)}] \\ \Re[\gamma^{(2)}] \\ \Im[\gamma^{(2)}] \label{eq:stmeq}
\end{pmatrix},
\end{equation}
where
\begin{eqnarray}
 \beta_{a}^{(n)}&=&\sum_i^{N_{H_a}}\cos({\bm G}\cdotp{\bm \tau}_i^{(n)}), \\
 \alpha_{a}^{(n)}&=&\sum_i^{N_{H_a}}\sin({\bm G}\cdotp{\bm \tau}_i^{(n)}), \\
 \beta_{c}^{(n)}&=&\sum_i^{N_{H_c}}\cos({\bm G}\cdotp{\bm \tau}_i^{(n)}), \\
 \alpha_{c}^{(n)}&=&\sum_i^{N_{H_c}}\sin({\bm G}\cdotp{\bm \tau}_i^{(n)}). 
\end{eqnarray}

In deriving \EQ{eq:stmeq} it has been taken into account that, since the potentials are real in real space, the real parts of $v_{H_\alpha}$ are symmetric while the imaginary components are antisymmetric in reciprocal space.

Using this methodology, passivant potentials were generated in this work for six semiconductor compounds (GaAs, AlAs, Si, Ge, CdSe and ZnO). For the extraction of the EPPs (the DFT calculations of $V_{\rm loc}({\bm r})$) a super-cell containing a slab of material in zincblende structure with surfaces on the (111) plane, and passivated with pseudohydrogens \cite{Huang2005,Zhang2016,Molina2012} was used. In this system, the atomic bonds linking the passivant to the bulk atoms are directed perpendicular to the surface plane, suggesting a high symmetry orientation. 

The procedure followed in each case was: (i) To find the optimal bond length for each pseudohydrogen by means of an structural relaxation. (ii) The DFT-LDA calculations of two slabs with surfaces on the (111) plane, with different number of atoms but with the same supercell size. (iii) The extraction of the EPPs by using \EQ{eq:stmeq}. In the step (iii), only the ${\bm G}$-vectors that correspond to the longest side of the supercell are used ([111] direction). This is done because the ${\bm G}$-point grid is the densest in this direction and contains all the information we need, including the potentials at small ${\bm G}$ values \cite{cardenas2012}. 

Special care has to be taken when evaluating \EQ{eq:stmeq}, since the $\gamma^{(i)}({\bm \tau},{\bm G})$ functions of \EQ{eq:gamma1} and \EQ{eq:gamma2} have been build from previously known AEPs, and they contain the error proper of the spherical approximation. This error is reflected in periodic nonuniformities in the passivant potentials in reciprocal space, which are located at ${\bm G}$-vectors that correspond to the bulk unit cell lattice vector, since the errors of the spherical approximation are placed on each atom position. Consequently, ${\bm G}$-points around an entire repetition of the two atom unit cell ${\bm G}$-vector have to be unregarded. The remaining points are interpolated to construct a continuous representation of the real and imaginary parts of the EPPs \cite{cardenas2012}, with no fitting or adjustment procedure of any kind.     

In \FG{fig1:paepCdSe} it is presented, as example, the real and imaginary parts of the EPPs for CdSe. Only the positive ${\bm G}$-points are shown, but the real components are symmetric and the imaginary components are antisymmetric. It is recognizable the typical shape of the AEPs \cite{cardenas2012} and that the imaginary parts, besides being smooth, are far from being negligible in comparison to the real components. This makes evident that the EPPs cannot be cast into the spherical approximation and, hence, the imaginary components need to be considered in calculations, otherwise the passivation would fail.

\begin{figure}
 \includegraphics[width=\figurewidth]{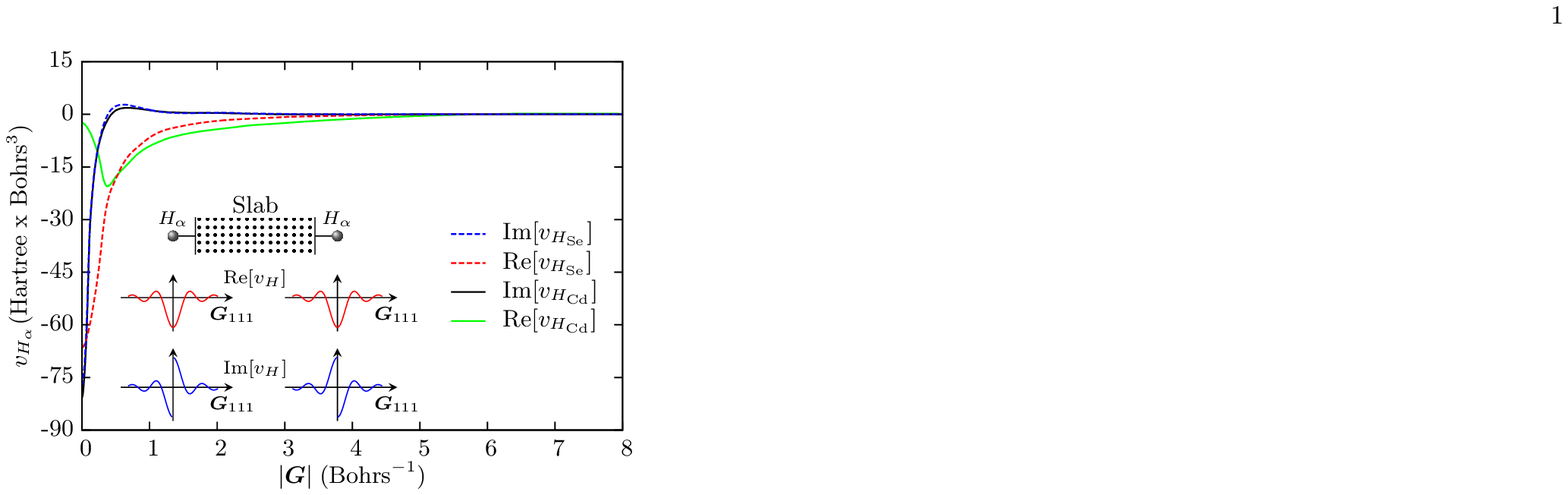}
\caption{Imaginary and real parts of the EPPs for Cd and Se along the [111] direction in CdSe zincblende structure. The diagram depicts how the imaginary component of the passivant potential is rotated depending on its location with respect to the surface.  \label{fig1:paepCdSe}}
\end{figure}


To implement the EPPs in further calculations, since the real components are spherically symmetric, they receive the same treatment of common bulk PSPs, they are then placed on the atom position and added up to the total potential. The imaginary components, to the contrary, are antisymmetric in ${\bm G}$-space and they need, therefore, to be somehow orientated according to the surface.    

Because of the symmetry, we can assume that the imaginary components of the potentials show their maximum amplitude along the passivant bond on the (111) surface, which is the [111] direction and the one chosen to generate the potentials in this work (\FG{fig1:paepCdSe}). The amplitude goes to zero when the ${\bm G}$-vector lies on the surface, and is lastly totally reflected when the ${\bm G}$-vector points inside the material perpendicular to the surface, as it is depicted in the diagram in figure \ref{fig1:paepCdSe}.     

Under this scheme, \EQ{eq:vlocstandard} now reads
\begin{equation} \label{eq:vlocvectors}
\begin{split}
V_{\rm loc} \argg &= \sum_\alpha^{N_{\rm species}}  \sum_{n}^{N_\alpha} 
 \e{-i {\bm G} \cdot {\bm \tau}_{\alpha n} } \Re[v_\alpha(|\bm G |)] \\ 
 & + i \sum_\beta^{N_{\rm H}} \sum_{m}^{N_\beta} \e{-i {\bm G} \cdot {\bm \tau}_{\beta m} }\Im[v_{\beta m}(|\bm G |)]\cos(\theta_{\beta m})  \quad  , 
\end{split}
\end{equation}
where $\alpha$ labels all the atomic species, including the passivants, $\beta$ runs only over the number of passivant types $N_{H}$, $\Im[v_{\beta m}(|\bm G |)]$ is the imaginary component of the passivants along the [111] direction (\FG{fig1:paepCdSe}) and $\theta_{\beta m}$ is the angle between the vector ${\bm p}_{\beta m}$ (normal to the surface and pointing outwards in the direction of each passivant $\beta_m$) and the actual ${\bm G}$-vector, whose cosine is defined by 
\begin{equation}
 \cos(\theta_{\beta m}) = \frac{{\bm p}_{\beta m}\cdotp{\bm G}}{|{\bm p}_{\beta m}||{\bm G}|} .
\end{equation}

All this means that for the integration of the passivants in the local potential of \EQ{eq:vlocvectors}, we need to define not only their positions, but also the normal vectors pointing to each one of the passivants.  

The generated EPPs were tested in calculations of slabs (of around five eight-atoms unit cell width) with surfaces on the (111), (110) and (100) planes and compared to DFT calculations. An ultimate comparison to DFT was done with a 68 atoms CdSe quantum wire (QWr) in wurtzite structure. In all cases the agreement is remarkable in both, the eigenvalues and wavefunctions, demonstrating the effectiveness of the non-spherical representation for passivants, the quality and high transferability of the EPPs, and the validity of the implementation method here presented.

In \FG{fig2:comparison}, the eigenvalues calculated with DFT \cite{gonze2002} and with the EPPs \cite{cardenas2012,Zirkelbach2015} are compared for the surfaces and slab systems mentioned before. In all cases the differences lie on the range of tens of meV. These results are remarkable, considering that the band gaps for bulk systems reach values as high as, for example, 2.5 eV for Si \cite{cardenas2012}. Besides, the error carried by the bulk potentials, due to the spherical approximation \cite{cardenas2012}, obviously affects the surface calculations done here, but the energy differences presented in \FG{fig2:comparison} don't evidence, even though, any quality reduction in comparison to bulk. 

\begin{figure}
 \includegraphics[width=\figurewidth]{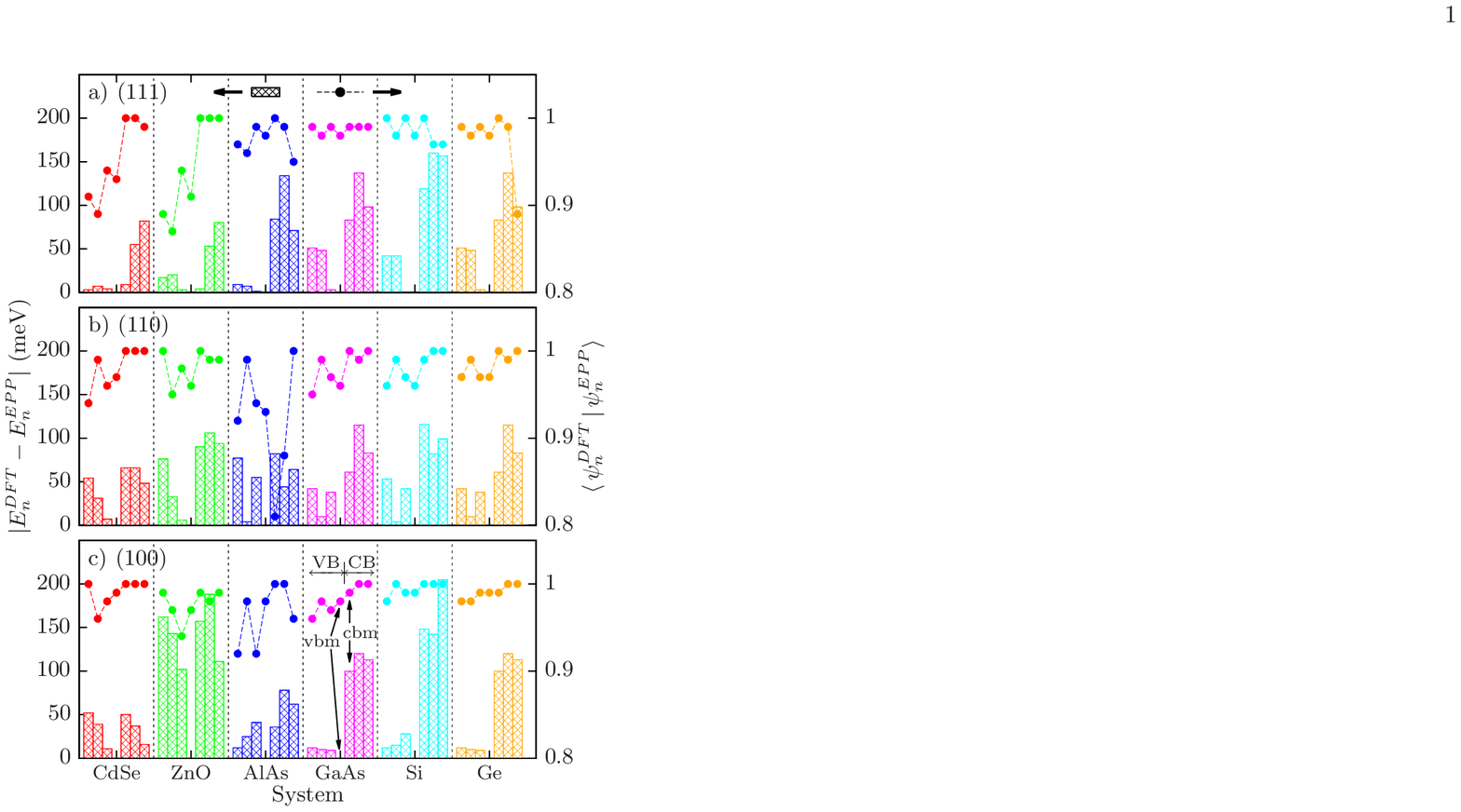}
\caption{Differences of the eigenvalues ($|E^{DFT}_n - E^{APP}_n|$ : left axes, bar symbols) obtained with DFT and with EPPs, referencing the vbm in both cases to the same value, and wavefunctions projection ($\bra{\psi^{DFT}_n}\psi^{EPP}_n\rangle$ : right axes dot symbols), for the slabs with surfaces along the a) (111), b) (110) and b) (100) planes, for the first four valence and first three conduction states, as indicated in the GaAs panel in (c). \label{fig2:comparison}}
\end{figure}

The effectiveness of the non-spherical EPPs is much better noticed when the EPPs wavefunctions are projected on the DFT wavefunctions ($\bra{\psi^{DFT}_n}\psi^{EPP}_n\rangle$). As shown in \FG{fig2:comparison} (right axes and dot symbols), the wavefunction products reveal most of the times an accuracy above the 99\% for all systems and directions. The lest accurate result is the cbm of AlAs, with an accuracy of 81\%, which keeps being a very good value. 

In order to further demonstrate the effectiveness of the EPPs, the veracity of the complex identity of the surface potentials and the method of their implementation into the atomistic calculation (\EQ{eq:vlocvectors}), a 68 atoms, including passivants, CdSe QWr in wurtzite structure has been studied. This is a very demanding test for the EPPs, specially in terms of transferability, since they are used in a different structure than the one used to generate them and that the surface doesn't lie on a particular plane. 

The results of the calculations done with the EPPs and DFT are shown in \FG{fig3:compQW}, where the differences in the eigenvalue and wavefunctions are listed for the firs five condition and valence states, as it was dine in \FG{fig2:comparison}. In the case of the eigenvalues, the difference never goes beyond 100 meV, a typical value for the bulk AEPs, showing that the EPPs don't introduce extra deviations in the calculations.    
\begin{figure}
 \includegraphics[width=\figurewidth]{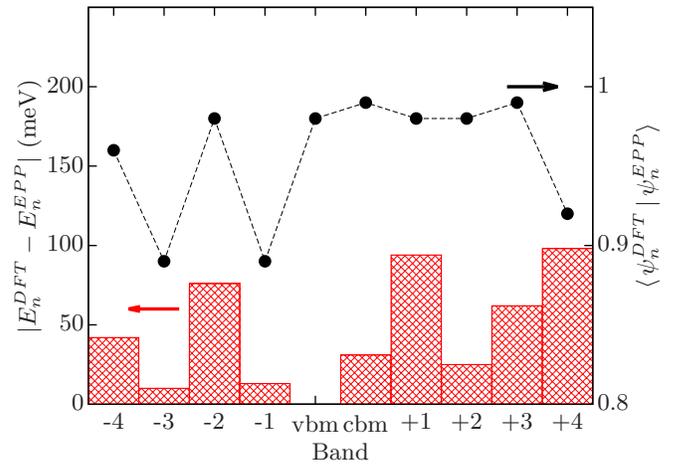}
\caption{Comparison of the eigenvalues (right axes, dot symbols) and wavefunctions (left axes, bar symbols) for the 68 atoms CdSe QWr, calculated with DFT and with EPPs, for the first five conduction and valence states. The eigenvalues have been aligned to the vbm.\label{fig3:compQW}}
\end{figure}

In a more striking manner than the agreement of the eigenvalues, the accuracy an reliability of the EPPs are better evaluated through the comparison of the wavefunctions. For the CdSe QWr, the results compare remarkably well to DFT, with an accuracy above 98\% for most of the cases, and above 89\% for all the results, as shown in \FG{fig3:compQW}. This level of accuracy has never been achieved before with any kind of effective of empirical potential in the spherical approximation.

\begin{figure}
 \includegraphics[width=\figurewidth]{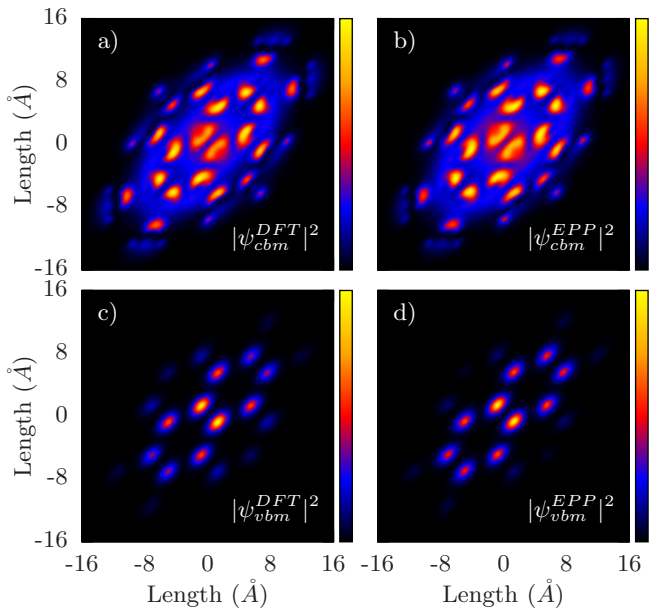}
\caption{Cbm wavefunctions as obtained with a) DFT ($\psi^{DFT}_{cbm}$) and with b) the EPPs ($\psi^{EPP}_{cbm}$), and of the vbm wavefunctions as obtained with c) DFT ($\psi^{DFT}_{vbm}$) and with d) the EPPs ($\psi^{EPP}_{vbm}$), for the CdSe wurtzite structure QWr. \label{fig4:CdSeQWr}}
\end{figure}

In figure \ref{fig4:CdSeQWr}, the cbm and vbm wavefunctions are compared graphically, where the average of the squared wavefunctions has been taken along the axial direction of the QWr. We can see how the EPPs wavefunctions reproduce with remarkable high fidelity the localization and fast oscillations of the wavefunctions found with DFT, warranting that the atomic character of the nanostructure is present in the EPPs calculations.     

As a final remark, it has to be mention that the similarity of the DFT results can be improved empirically by tuning the potentials or by adding weights, as done with alloys \cite{cardenas2012}, on surfaces where the passivant species mixe, like the (110) one, or where bonds open to capture to passivants, like in the (100) plane. But this it not done here, to keep the approach as {\it ab inito} and free of fittings as possible.


To summarize, a new type of effective passivant pseudopotential has been presented. It has been shown that the spherical approximation is not suitable for describing passivants and that, instead, for the proper consideration of the surface effects, the passivant potentials have the be included as complex quantities in reciprocal space. This requires the definition of the passivant orientation over the surface of the nanostructure and demands changes in the way the atomic passivant potentials are implemented in the calculations. Both, the derivation and implementation methods have been also described here. 

The accuracy and reliability of the method have been demonstrated by comparison to DFT calculations of slabs with surfaces along the (111), (110), (100) planes for six different semiconductor compounds. Besides, a 68 CdSe QWr in wurtzite structure has been studied in the same manner. In all cases the results compare remarkably well to DFT, demonstrating the validity of the methodology and the high quality of the 
potentials in therms of accuracy and transferability. 

Concerning practicality, the methodology presented not only allows for the generation of high quality potentials, but is also free of any fitting or extra tedious tuning procedure. On the other hand, the idea of describing passivant pseudopotentials as complex quantities is completely general and can be applied to any method, such as AEPs or EPM.

\begin{acknowledgements}
I would like to thank Professor Ricardo Vega, director of research group FITES of the {\it Universidad del Atl\'antico}, for providing the computational tools and the space for the development of this work.
\end{acknowledgements}

\end{document}